\documentclass{appolb}
\usepackage{graphicx}
\usepackage{amsmath}

\begin{document}

\title{Density Functional Theory Approach to Nuclear Fission
\thanks{Presented at the XLVII${\text{th}}$ Zakopane Conference on Nuclear Physics}
}

\author{N. Schunck
\address{Physics Division, Lawrence Livermore National Laboratory, CA 94551, USA, }
}

\maketitle

\newcommand{\gras}[1]{\boldsymbol{#1}}

\begin{abstract}
The Skyrme nuclear energy density functional theory (DFT) is used to model 
neutron-induced fission in actinides. This paper focuses on the numerical 
implementation of the theory. In particular, it reports recent advances in 
DFT code development on leadership class computers, and presents a detailed 
analysis of the numerical accuracy of DFT solvers for near-scission calculations. 
\end{abstract}


\section{Introduction}

In spite of successful applications in energy production and national security, 
relatively little is known about the fundamental mechanisms of the nuclear fission 
process. Even today, the most widely-used theories of fission rely on the 
macroscopic-microscopic approach to nuclear structure and semi-classical dynamics 
based on the Langevin equations. These methods are quite powerful but, in the 
long term, cannot be expected to yield the predictive power needed to understand 
the fission of very neutron-rich nuclei in the fission-recycling stage of the 
formation of elements, or to give enough accuracy for precise simulations of new 
generations of nuclear reactors.

Already in the 1980ies, promising attempts were made to understand fission in a 
microscopic framework based on the self-consistent nuclear mean-field theory with 
effective pseudo-potentials \cite{[Ber84],[Ber89]}. At the time, the computing 
power was not sufficient for these approaches to compete with more empirical 
models, but these pioneer works yielded a lot of insight on the quantum mechanics 
of fission. In the recent years, the rapid development of leadership-class 
computers scaling to hundreds of thousands, and soon millions, of processing units 
has given us for the first time the computing power needed to successfully 
implement the full microscopic theory of fission. In parallel, novel forms of 
scientific collaborations gathering nuclear theorists, applied mathematicians and 
computer scientists have considerably improved our ability to utilize such 
large-scale machines \cite{unedf}.

The goal of this paper is to provide some tools to verify and validate DFT 
simulations of nuclear fission. In particular, we pay special attention to the 
problem of the numerical accuracy of DFT solvers at very large deformations. 
After a brief reminder on the theoretical framework, we recall some of the 
recent developments in DFT calculations on leadership class computers, then 
give a detailed analysis of truncation errors arising in DFT implementations 
using the one-center finite harmonic oscillator basis. 

\section{Theoretical Framework}

The nucleus is described in the local density approximation at the Hartree-Fock 
Bogoliubov (HFB) approximation. The particle-hole channel is modeled by effective 
pseudo-potentials up to second-order derivatives in the density. In practice, 
this is equivalent to using zero-range effective interactions of the Skyrme 
type \cite{[Dob95],[Dob96b]}. The results presented below were thus obtained with 
the SkM* parameterization of the Skyrme interaction \cite{[Bar82]}. The 
particle-particle channel is characterized by a density-dependent contact 
interaction with mixed volume and surface character \cite{[Dob02]}. An energy 
cut-off of $E_{\text{cut}} = 60$ MeV is used to reduce the number of quasi-
particles in the definition of the densities. The HFB equations are solved in a 
one-center harmonic oscillator basis.

In the DFT picture of nuclear fission, the HFB energy of the nucleus depends on 
an ensemble of collective variables $\gras{q} = (q_{1}, \dots, q_{N})$. These can 
be, for example, variables describing the nuclear shape, excitation or spin. In  
this work, we considered as collective variables the expectation value (on the 
HFB ground-state) of the multipole moments $\hat{Q}_{\lambda\mu}$. In practice, 
the axial $\hat{Q}_{20}$ and triaxial $\hat{Q}_{22}$, as well as the mass octupole 
$\hat{Q}_{30}$ and hexadecapole $\hat{Q}_{40}$ moments were considered. The 
collective space is thus four-dimensional. Expectation values of $\hat{Q}_{\lambda\mu}$ 
will simply be denoted $Q_{\lambda\mu} \equiv \langle \hat{Q}_{\lambda\mu}\rangle$.

In the actinide region, the part of the potential energy surface relevant to 
nuclear fission spans a rather large range in deformations. The axial quadrupole 
moments runs typically from $\sim 30$ b in the ground-state to nearly 600 b at 
scission for symmetric fission; the octupole moment from 0 to about 70 b$^{3/2}$ 
for very asymmetric fission (cluster radioactivity, see \cite{[War11],[War12]}); 
the hexadecapole moment from nearly $\sim 3$ b$^{2}$ near the ground-state to 
typically more than $\sim 350$ b$^{2}$ for symmetric fission. Assuming for sake of 
simplicity a uniform sampling of each degree of freedom, and a $1$ b$^{\lambda/2}$ 
mesh size, the size of the collective space is more than 1.4 millions points, only 
for the axial collective variables. Adding triaxiality multiplies this estimate by 
another 2 orders of magnitude. 

\section{Large-Scale Potential Energy Surfaces}

In this section, we present the performance of our DFT solver, and provide a 
detailed analysis of convergence properties of our Skyrme HFB calculations in 
the case of $^{240}$Pu.

\subsection{DFT Solvers on Leadership Class Computers}

All calculations were performed with the DFT solvers HFBTHO \cite{[Sto12]} 
and HFODD \cite{[Sch11]}. Both codes solve the HFB equations in the harmonic 
oscillator (HO) basis. The program HFBTHO assumes axial and time-reversal 
symmetry, while HFODD is fully symmetry-unrestricted. The two programs have 
been benchmarked against one another and agree within a few eV for an axial 
configuration \cite{[Sto12]} . 

\begin{figure}[htb]
\begin{center}
\includegraphics[width=0.8\linewidth]{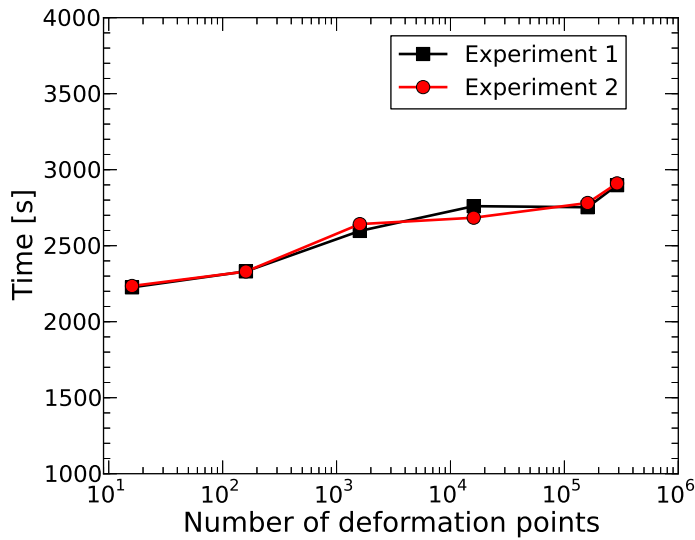}
\caption{Performance of the DFT solver HFODD on the Titan supercomputer at the 
Oak Ridge Leadership Computing Facility in Oak Ridge. Each experiment measures 
the time of performing six full HFB iterations. The term 'deformation point' 
refers to a set of constraints on multipole moments, i.e. a point in the 
collective space.}
\label{fig:titan_scaling}
\end{center}
\end{figure}

Owing to the block-diagonal structure of the HFB matrix induced by axial 
symmetry, the typical runtime of HFBTHO is of the order of the minute 
(depending on the type of nucleus, size of the basis and 'difficulty' of 
converging the HFB iterations). By contrast, it is of the order of several 
hours for HFODD. In practice, HFBTHO is used as pre-conditioner for HFODD: 
for any given point $\gras{q}$ in the collective space, the HFB iterations 
are first solved with HFBTHO, and the densities at convergence are used to 
initialize HFODD. If the point of the collective space is axial, no additional 
iterations are therefore needed.

Clearly, the very large size of the collective space together with the current 
runtime of the DFT solvers requires using today's most powerful supercomputers. 
A lot of effort was, therefore, devoted to porting our codes to leadership 
class computers, and ensuring that good scaling with the number of processing 
units could be achieved. From a computational point of view, mapping the nuclear 
collective space in DFT is a naturally parallel problem. The code HFODD has, 
therefore, a hybrid MPI/OpenMP programming model, where points in the collective 
space are distributed across the MPI grid, and on-node multi-threading enables 
to take advantage of highly optimized linear algebra libraries. Figure 
\ref{fig:titan_scaling} shows the performance of HFODD on the Titan supercomputer 
at the Oak Ridge Leadership Computing Facility. In this experiment, up to 300,000 
processors were used in parallel. The slight degradation of the computing time 
is due to the original input/output backend, which has not been optimized and 
taxes the operating system at large scale.

\subsection{Numerical Accuracy}

Modeling nuclear fission requires to explore regions of the collective 
space with extreme deformations. The finite size of the HO basis may thus lead 
to a significant dependence of the results on the basis parameters. In our 
calculations, we only considered axially-deformed bases, characterized by 
a quadrupole deformation $\beta$ and an oscillator frequency $\omega_{0}$. 
The maximum number of shells is denoted by $N_{\text{max}}$ and the maximum 
number of states by $N_{\text{states}}$. In the case of a full spherical HO 
basis, the two are related through the well-known relation 
$N_{\text{states}} = (N_{\text{max}}+1)(N_{\text{max}}+2)(N_{\text{max}}+3)/6$. 
This relation is not valid anymore for a deformed basis. In practice, we 
introduce two cut-offs, one on the number of shells and another on the number 
of states. Note that the cut-off on the number of states is only the practical 
consequence of using a symmetry-unrestricted solver, for which the size of 
the matrices involved goes approximately as $2N_{\text{max}}^{3}$. By contrast, 
the block structure induced by the built-in axial symmetry in HFBTHO would 
enable to consider all full shells up to $N_{\text{max}}$.

Basis dependence of the calculations thus comes from 4 parameters (i) the 
maximum number of shells $N_{\text{max}}$, (ii) the maximum number of states 
$N_{\text{states}}$, (iii) the oscillator frequency $\omega$ and (iii) the 
basis deformation $\beta_{2}$. In principle, at every point $\gras{q}$ in 
the collective space, we should seek the HFB solution that is the minimum in 
this 4-dimensional parameter space. Clearly, such a strategy is not sustainable 
even on today's largest computers. Instead, one is bound to estimate truncation 
errors by exploring the parameter space locally, and extracting asymptotic expressions.
 
\begin{figure}[!sht]
\center
\includegraphics[width=0.8\linewidth]{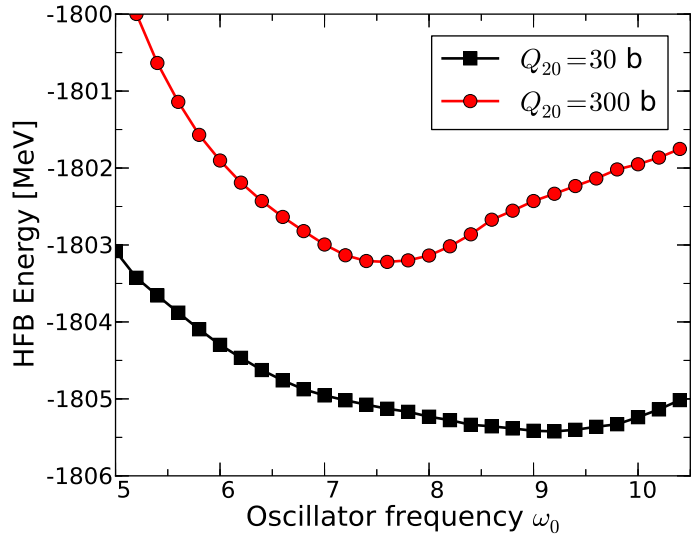}
\caption{(color online) Convergence of the HFB energy as function 
of the oscillator frequency $\omega_{0}$, for two configurations characterized 
by $Q_{20} = 30$ b (black squares) and $Q_{20}=300$ b and $Q_{40} = 120$ 
b$^{2}$ (red circles). The deformation $\beta$ is adjusted according to the 
formula (\ref{eq:beta}).
}
\label{fig:HObasis1}
\end{figure}

As a first example, we show in figure \ref{fig:HObasis1} the dependence of 
the HFB energy on the spherical-equivalent frequency of the harmonic oscillator 
$\omega_{0}$, that is the frequency such that $\omega_{0}^{3} = 
\omega_{x}\omega_{y}\omega_{z}$. The deformation of the basis is fixed at 
each point according to the formula (\ref{eq:beta}). Two typical points in 
the collective space are considered, one near the ground-state with deformation 
$Q_{20} = 30$ b, one way past the second barrier on the descent to scission 
at $Q_{20} = 300$ b and $Q_{40} = 120$ b$^{2}$. We note that the dependence 
on $\omega_{0}$ is more marked at large deformations, and that the optimal 
frequency shifts toward smaller values as the deformation increases, which 
is consistent with the need to then include basis states with a larger 
spatial extension. Importantly, it is possible to extend this analysis and 
extract an empirical fit $\omega_{0}(Q_{20})$ giving the optimal basis frequency 
as function of the quadrupole moment of the collective point. In our tests, we 
found that the expression
\begin{equation}
\omega_{0} = 
\left\{ \begin{array}{l}
0.1 \times Q_{20}e^{-0.02 Q_{20}} + 6.5\; \text{MeV}\ \text{if}\ |Q_{20}| \leq 30\;  \text{b} \\
8.14\;  \text{MeV} \ \text{if}\ |Q_{20}| > 30\;  \text{b}
\end{array}
\right.
\label{eq:omega}
\end{equation}
gives a reasonably accurate fit of the frequency as function of the quadrupole 
moment.

\begin{figure}[!sht]
\center
\includegraphics[width=0.8\linewidth]{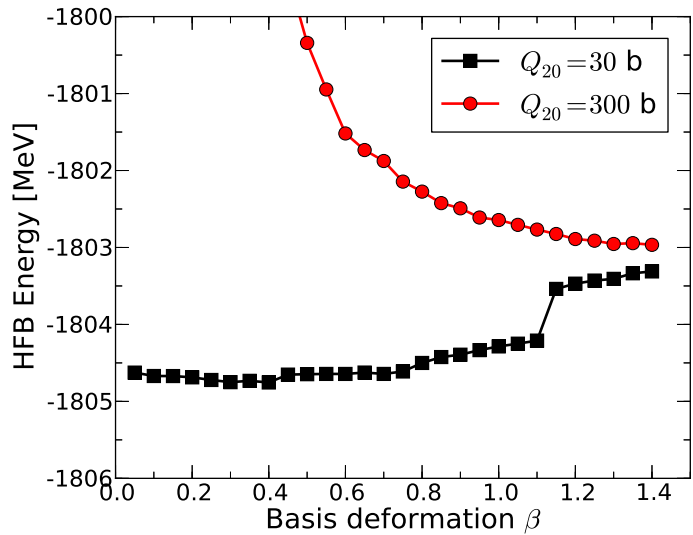}
\caption{(color online) Convergence of the HFB energy as function of the basis 
deformation $\beta$. The oscillator frequency is adjusted according to the 
formula (\ref{eq:omega}). For the second configuration, an additional 
constraint on $\hat{Q}_{60}$ to $Q_{60} = 150$ b$^{3}$ was added.
}
\label{fig:HObasis2}
\end{figure}

The dependence of the HFB energy on the deformation of the basis, for a fixed 
$N_{\text{max}}$ and basis deformation, is illustrated in figure 
\ref{fig:HObasis2}. Not surprisingly, the minimum is always obtained for basis 
deformation that are 'close' to the requested value of the axial quadrupole 
moment. Note that the dependence on deformation is rather marked. However, as for 
the oscillator frequency, it is {\it a priori} possible to obtain a fit 
$\beta(Q_{20})$ such that the optimal basis deformation is chosen at point in the 
collective space. Our tests showed that the simple formula
\begin{equation}
\beta = 0.05\sqrt{Q_{20}}
\label{eq:beta}
\end{equation}
provides a reasonable expression that remains applicable up to the largest 
values of $\hat{Q}_{20}$.

\begin{figure}[!ht]
\center
\includegraphics[width=0.8\linewidth]{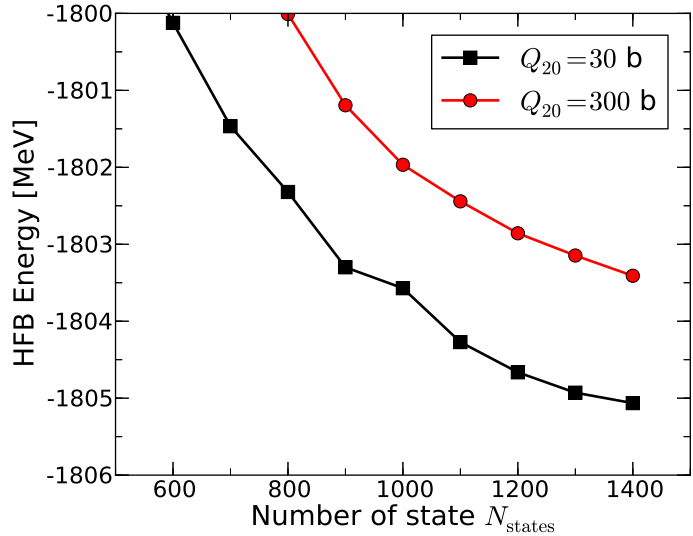}
\caption{(color online) Convergence of the HFB energy as function of the number 
of states $N_{\text{states}}$. 
}
\label{fig:HObasis3}
\end{figure}

Last but not least, we show in figure \ref{fig:HObasis3} the error induced 
by the truncation of the number of states. Contrary to the previous two 
parameters of the HO basis, this truncation is imposed by memory limitations, 
and can not really be mitigated: for a given value of $N_{\text{max}}$, the 
optimal number of states is always given by
$N_{\text{states}} = (N_{\text{max}}+1)(N_{\text{max}}+2)(N_{\text{max}}+3)/6$, 
a number that can grow very large for large $N_{\text{max}}$ For example, at 
$N_{\text{max}} = 30$, we have $N_{\text{states}} = 5456$. Taking into account 
the spin degree of freedom, the total number of basis states is more than 10,000, 
which implies that the size of the HFB matrix exceeds 20,000 $\times$ 20,000. At 
this time, it is not possible to handle in a reasonable time frame iterative 
processes involving dense, complex matrices in double precision of that size. As 
can be seen from figure \ref{fig:HObasis3}, restricting $N_{\text{states}}$ to 
manageable values around $N_{\text{states}}\approx 1000$-$1200$ may easily lead 
to 2 to 3 MeV errors beyond the second fission barrier.

To finish this section, we would like to emphasize two important consequences of 
basis truncation effects:
\begin{itemize}
\item At constant truncation, the error increases with deformation, albeit not 
necessarily linearly. This is bound to have a very significant impact for, e.g., 
calculations of barrier penetrability, since the numerical precision is not the 
same at the entry and exit points, and errors of the order of the MeV can lead to 
orders of magnitude uncertainties on fission lifetimes;
\item Truncations magnify the impact of discontinuities in the potential energy 
landscape. In practice, calculations with very different basis characteristics 
initialized with similar density/wave-functions could converge to two 
very different points of the multi-dimensional PES. This effect is the reason 
why, in the lower panel of figure \ref{fig:HObasis2}, an additional constraint 
on $\hat{Q}_{60}$ had to be added: without it the calculation did not converge to 
the same point in the collective space at small and large basis deformations.
\end{itemize}

\section{Conclusions}

The nuclear energy density functional theory is currently the only viable option 
to achieve a microscopic description of nuclear fission. On-going development 
of leadership class computing facilities all over the world offer a unique 
opportunity to finally develop the nuclear DFT at very high precision. In this 
work, we have discussed some of the numerical uncertainties associated with 
implementations of DFT in the one-center harmonic oscillator basis. They clearly 
point to the need of developing bases that are better adapted to the extreme 
elongations characterizing the region near scission.

\section{Acknowledgments}
\label{sec:acknowledgments}

Discussions with W. Younes are warmly acknowledged. Support for this work 
was provided in part through Scientific Discovery through Advanced Computing 
(SciDAC) program funded by U.S. Department of Energy, Office of Science, 
Advanced Scientific Computing Research and Nuclear Physics. It was partly 
performed under the auspices of the US Department of Energy by the Lawrence 
Livermore National Laboratory under Contract DE-AC52-07NA27344. Funding was 
also provided by the United States Department of Energy Office of Science, 
Nuclear Physics Program pursuant to Contract DE-AC52-07NA27344 Clause 
B-9999, Clause H-9999 and the American Recovery and Reinvestment Act, Pub. 
L. 111-5. An award of computer time was provided by the Innovative and Novel 
Computational Impact on Theory and Experiment (INCITE) program. This 
research used resources of the Oak Ridge Leadership Computing Facility 
located in the Oak Ridge National Laboratory, which is supported by the 
Office of Science of the Department of Energy under Contract DE-AC05-00OR22725. 
It also used resources of the National Energy Research Scientific Computing 
Center, which is supported by the Office of Science of the U.S. Department 
of Energy under Contract No. DE-AC02-05CH11231.

\bibliographystyle{unsrt}
\bibliography{zakopane}

\end{document}